\pdfoutput=1 
\documentclass{IEEEtran}
\usepackage{cite}
\usepackage{amsmath,amssymb,amsfonts}
\usepackage{algorithmic}
\usepackage{graphicx}
\usepackage{textcomp}
\usepackage{multirow}
\usepackage{hhline}
\usepackage{subfigure}
\usepackage{boldline}
\usepackage{amsmath}
\usepackage{color}
\usepackage{gensymb}
\usepackage{fancyhdr}
\IEEEoverridecommandlockouts
\def\BibTeX{{\rm B\kern-.05em{\sc i\kern-.025em b}\kern-.08em
    T\kern-.1667em\lower.7ex\hbox{E}\kern-.125emX}}

\begin{document}

\title{{\fontsize{24}{26}\selectfont{ Attenuation of Several Common Building Materials in Millimeter-Wave Frequency Bands: $28$, $73$ and $91$ GHz }}\break\fontsize{16}{18}\selectfont
}
\author{Nozhan Hosseini, \IEEEmembership{Member, IEEE}, Mahfuza Khatun, \IEEEmembership{Member, IEEE},  Changyu Guo,
\IEEEmembership {Student Member, IEEE},  Kairui Du, \IEEEmembership{Member, IEEE}, Ozgur Ozdemir, \IEEEmembership{Member, IEEE}, David W. Matolak, \IEEEmembership{Senior Member, IEEE}, Ismail Guvenc \IEEEmembership{Senior Member, IEEE}, Hani Mehrpouyan, \IEEEmembership{Member, IEEE}
  \thanks{* This work was supported by NASA, under award number NNX17AJ94A}
}

\maketitle
  
\begin{abstract}
 Future cellular systems will make use of millimeter wave (mmWave) frequency bands. Many users in these bands are located indoors, i.e., inside buildings, homes, and offices. Typical building material attenuations in these high frequency ranges are of interest for link budget calculations. In this paper, we report on a collaborative measurement campaign to find the attenuation of several typical building materials in three potential mmWave bands ($28$, $73$, $91$ GHz). Using directional antennas, we took multiple measurements at multiple locations using narrow-band and wide-band signals, and averaged out residual small-scale fading effects. Materials include clear glass, drywall (plasterboard), plywood, acoustic ceiling tile, and cinder blocks. Specific attenuations range from approximately 0.5 dB/cm for ceiling tile at 28 GHz to approximately 19 dB/cm for clear glass at 91 GHz.
\end{abstract}


\section{Introduction}
\label{sec:intro}
\begin{figure}[!t]
\centerline{\includegraphics[width=\columnwidth]{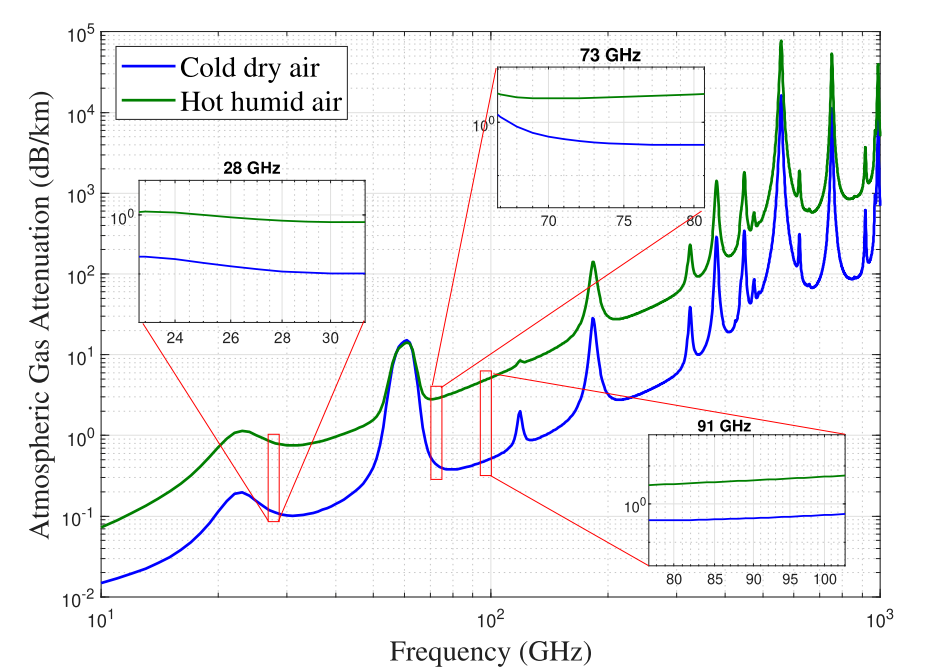}}
\caption{Atmospheric gas attenuation versus frequency.}
\label{fig_athmos}
\end{figure}

To achieve higher throughput in future generations of wireless communication systems, e.g., $20$ Gbit/s download speed in $5$G, mmWave frequency bands are of interest.  Fig. \ref{fig_athmos} depicts the well known atmospheric attenuation versus frequency for frequencies $10-1000$ GHz \cite{ITU-840}. As can be seen, ranges between $10$-$40$ and $70$-$100$ GHz have lower attenuation than adjacent bands. In this paper, we investigate center frequencies in these ranges, specifically at $28$, $73$ and $91$ GHz. Fig. \ref{fig_athmos} shows two plots of atmospheric gas attenuation for ``cold and dry" and ``hot and humid air," for water vapor density equal to 7.5  $g/m^3$ and 50   $g/m^3$, respectively,  based on \cite{ITU-840}, \cite{ITU-836}. Note that the value of 7.5 $g/m^3$ is the average value for $50\%$ of the time in dry areas on the earth, and 50 $g/m^3$ is the maximum value for $10\%$ of the time in the most humid parts of the earth.

\subsection{Literature Review}

There have only been a few studies for  attenuation of different construction materials at mmWave frequencies. This attenuation is also sometimes termed ``penetration loss," particularly when the propagation is from outdoor to indoor (or vice-versa).
In one notable work,  \cite{rappa_globcom19} the authors studied reflection loss, scattering and the loss of partition structures (partition loss) at three mmWave frequencies\textemdash $28$, $73$ and $140$ GHz. As expected the authors found larger partition loss at higher frequencies than in lower bands. We note that the amount of attenuation can also strongly depend on the composition of the materials and antenna polarization. These results are limited to clear glass and drywall building materials \cite{rappa_globcom19}.

In another study from the same research group, the authors performed  reflection and penetration loss measurements of common building materials in dense urban environments in New York City at $28$ GHz \cite{zhao201328,rqppaport2013Itwillwork}. They found that indoor-to-outdoor attenuation through the building materials is larger than that of indoor-to-indoor and outdoor-to-outdoor propagation. Results show that mmWave signals can penetrate well through several indoor materials and can incur strong reflections from the external building materials outdoors. In addition to that work, in ~\cite{ryan2017indoor} the authors observed that penetration loss at $73$ GHz does not necessarily increase or decrease based on the antenna polarization; this measurement campaign was performed in a typical indoor office environment. 

Another study on  propagation path loss in a building at $60$ GHz was conducted in~\cite{1296643}. These results showed that at this frequency, because of the very large penetration loss, the signal can be effectively confined to a single room. They also found very low RMS delay spreads from multipath components created by reflectors within a single room. In ~\cite{mahfuza_globcom19}, we measured penetration loss of building materials at two mmWave bands \textemdash$73$ GHz and $81$ GHz \textemdash on the campus of Boise State University and showed that outdoor building materials can have higher attenuation than the indoor materials due to the multi-layer structure and larger thickness of these outdoor materials.

The authors in \cite{usc1} reported measured building penetration losses in a suburban neighborhood for a 3GPP model at 28 GHz. Three example houses were investigated. They reported 9 dB median building penetration loss for a home and plain-glass windows, and 15 dB for a renovated home with low emissivity windows, and 17 dB for a new construction that has foil backed insulation and low emissivity windows. Obviously, their measurement campaign only considered a specific outdoor-to-indoor scenario and not precise material loss quantification. 

In \cite{3_to_24ghz_paper}, the authors reported on a penetration loss measurement campaign to examine window penetration loss and building entry loss of a traditional office building from 3.5 to 24 GHz. They chose a traditional office building (on the Chosun University campus), where the exterior walls were reinforced concrete and glass windows are double-glazed glass. They provided results using building entry loss for different receiver locations. They reported that loss increased as the receiver (Rx) was located more deeply within the building. At all designated
locations inside the building, the loss at 24 GHz was lower than that at 18 GHz, and loss values at 6 GHz were actually greater than that at 10 or 24 GHz. Such a measurement can be considered a multi-material test that cannot easily be generalized to other buildings.

 \begin{figure}[!t]
\centerline{\includegraphics[width=\columnwidth]{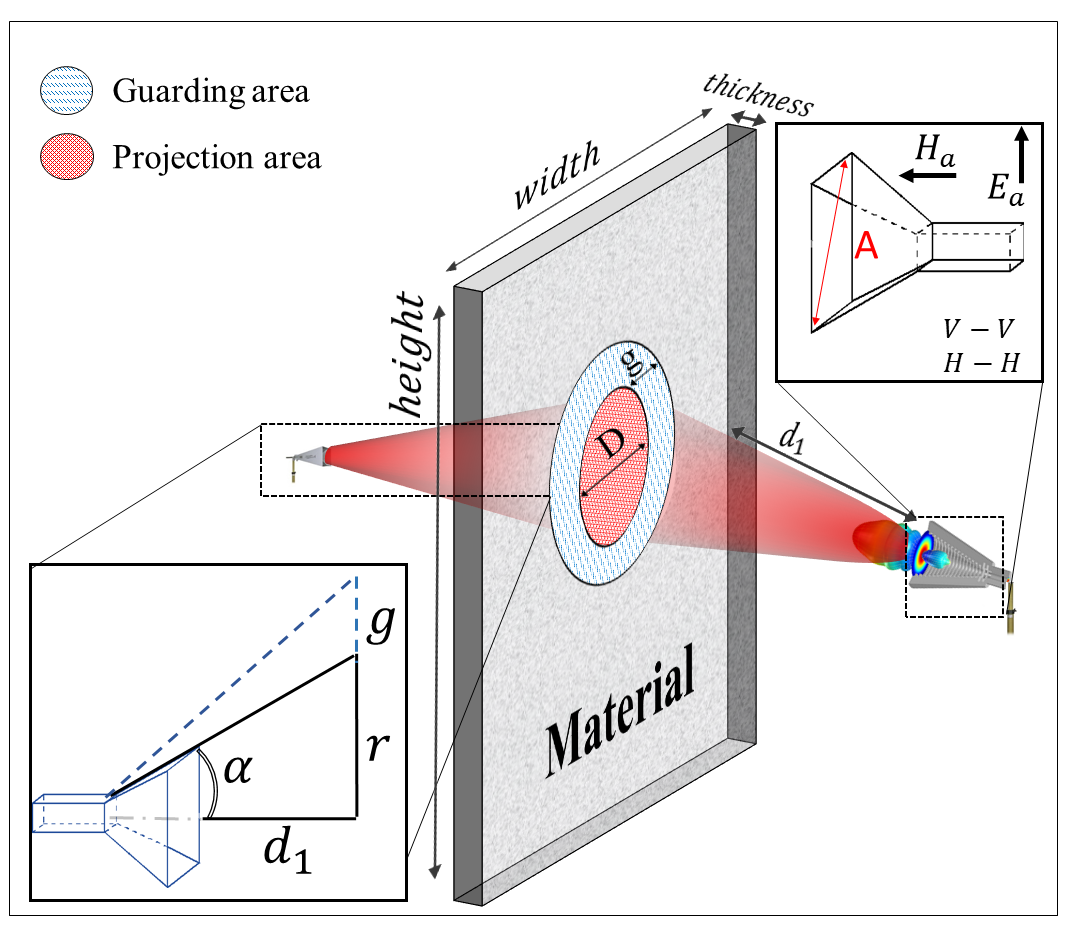}}
\caption{Geometry for determining projection area and material dimensions.}
\label{setup}
\end{figure}
 \begin{figure}[!t]
\centerline{\includegraphics[width=\columnwidth]{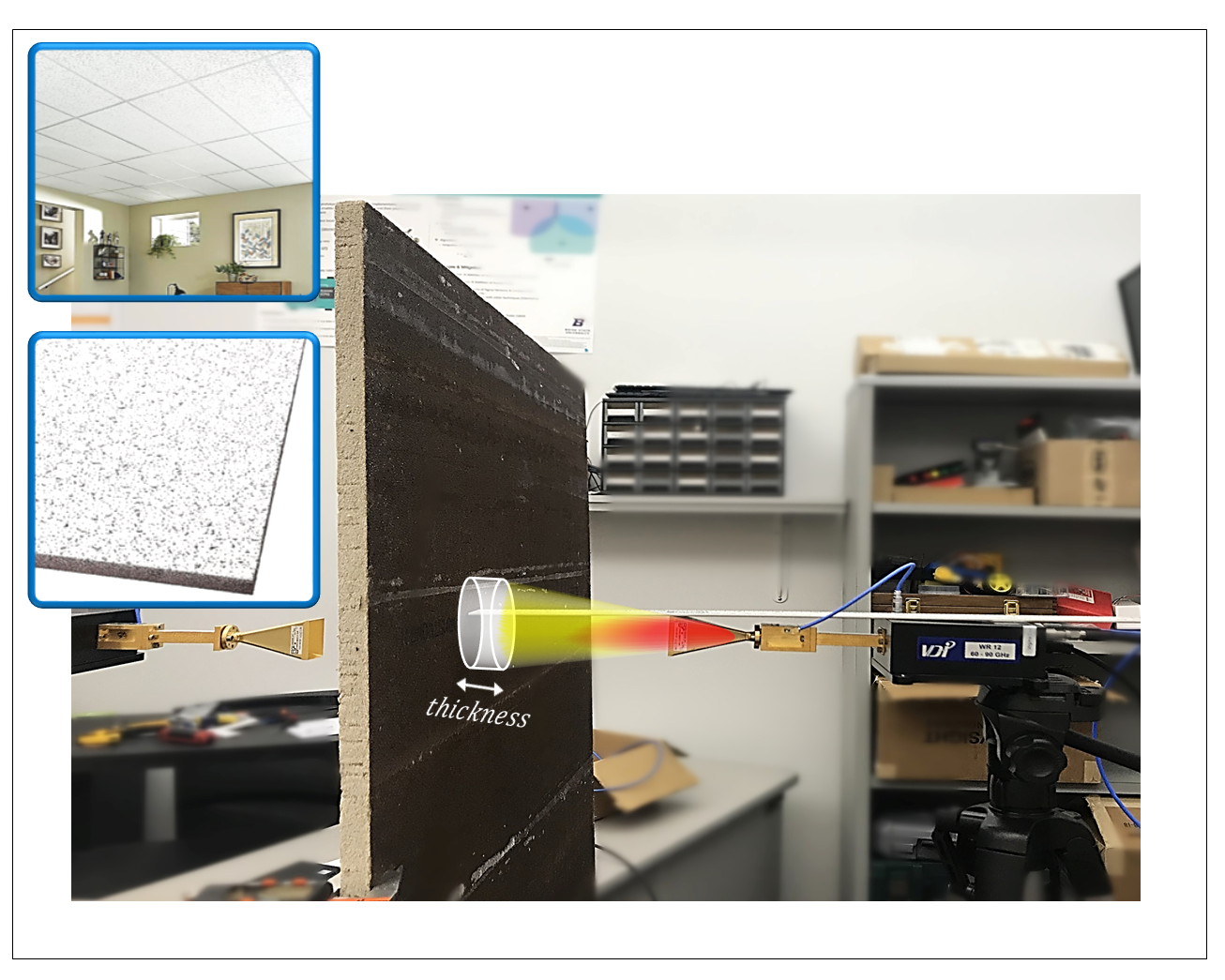}}
\caption{One snapshot of real measurement setup at 73 GHz}
\label{setup2}
\end{figure}
 In lower frequency bands, the authors of \cite{finland_paper_low_freq} defined three measurement methods termed outdoor-to-indoor, far-field, and near-field penetration loss, and compared these results. They claimed that the three methods gave similar penetration loss results. Specifically, their materials were three-layer window glass and a brick wall, and their frequency range was 1 to 17 GHz. They found penetration losses of 10-20 dB between 5 and 9 GHz and approximately 30-50 dB for frequencies above 7 GHz for the brick wall. For their three layer window, they stated that the loss increased to 20 dB for frequencies up to 4 GHz and decreased again below 10 dB at 9 GHz, hence observing the window acts a band pass filter. This result is based on specific materials in specific locations and losses were not quantified in dB as a function of material depth.  

In a different approach, the authors in \cite{three_layer_absorber} constructed a three-layer wireless LAN absorber with building materials that had about $15$ dB attenuation in two wireless LAN frequency ranges ($2.4-2.5$ GHz and $5.15-5.25$ GHz). The best absorber they composed had three layers of fiber reinforced cement board of thickness $3.1$ mm, an air layer of width $3.2$ mm, and a third layer of fiber reinforced cement board of thickness $29.3$ mm. 

Another comprehensive study was done in \cite{different_concrete} to investigate the attenuation of several different types of concrete building  samples. The authors used samples at different temperature, humidity, age and water-to-cement ratios. They concluded that concrete blocks, which may have water-to-cement ratio of 1, have significant variation in their RF attenuation. At higher frequencies around $16$ to $18$ GHz, they show that two different samples have attenuations that differ by nearly 50 dB. Therefore, no universal attenuation value can be claimed for these concrete blocks.
\begin{table}[]
\centering
\caption{Material List}
\label{tab:material_model}
\begin{tabular}{ V{3}c||c ||cV{3}  }
\hlineB{3}
     \textbf{Material}    & {\textbf{Manufacturer}} & {\textbf{Dimension}}  \\    \hlineB{3}
    
       Plywood     &   \begin{tabular}[c]{@{}l@{}}Plytanium $15/32$\\  CAT PS1-09, Pine\end{tabular}  & $4 ft$ $\times 8 ft$ \\ 
      \hline
      Acoustic     & \begin{tabular}[c]{@{}l@{}}Armstrong Acoustic\\   Panel Ceiling Tiles\end{tabular}   &  $48 in$ $\times 24 in$  \\ 
      \hline
      Clear Glass &    Gardner Glass Products & $30 in$ $\times 36 in$   \\ 
      \hline
     Drywall     &   \begin{tabular}[c]{@{}l@{}}ToughRock Fireguard\\   Drywall Panel \end{tabular}   & $5/8$-in $\times 4 ft$  $\times 8 ft$\\ 
     \hline
     Concrete block    & \begin{tabular}[c]{@{}l@{}}Standard Core\\ Concrete Block  \end{tabular}   &  $8$-in $\times  8 in$ $\times 16 in$   \\

\hlineB{3}
\end{tabular}
\end{table}

\begin{table*}[!ht]
	\centering
	\caption{Hardware Specifications of the $28$, $73$ and $91$ GHz Attenuation Measurements }
	\label{tab:hardware_spe}
	\begin{tabular}{|p{150pt}||p{50pt}|p{50pt}|p{50pt}|}  
			\hhline{-|-|--}
		
		\centering \textbf {Campaign}
		&\centering   \textbf {NCSU}  & \centering \textbf{BSU} &
		
		 \textbf{\textcolor{white}{......} USC} \\ 
		\hhline{=::=:==}
		
		\multicolumn{1}{|c||}{	Center frequency (GHz)} & \multicolumn{1}{c|}{28 } & \multicolumn{1}{c|}{73 } &  \multicolumn{1}{c|}{91 } \\ 
		\hhline{-|-|--}
		
				\multicolumn{1}{|c||}{	Wave Length (mm)} & \multicolumn{1}{c|}{10.7 } & \multicolumn{1}{c|}{4.1 } &  \multicolumn{1}{c|}{3.29 } \\ 
		\hhline{-|-|--}

		\multicolumn{1}{|c||}{Wideband signal } &   \multicolumn{1}{c|}{Zadoff-Chu (ZC)} & \multicolumn{1}{c|}{Chirp} &  \multicolumn{1}{c|}{Chirp} \\ 
		\hhline{-|-|--}

		\multicolumn{1}{|c||}{Bandwidth (MHz) } &   \multicolumn{1}{c|}{1500} & \multicolumn{1}{c|}{500} &  \multicolumn{1}{c|}{500 } \\ 
		\hhline{-|-|--}
		
		\multicolumn{1}{|c||}{Narrowband signal } &   \multicolumn{1}{c|}{ Not applicable}&   \multicolumn{1}{c|}{ CW}&   \multicolumn{1}{c|}{ CW} \\ 
		\hhline{-|-|--}
		
		\multicolumn{1}{|c||}{TX and RX antenna Type} &  \multicolumn{3}{c|}{ Rectangular Horn Antenna}  \\ 
		\hhline{-|-|--}

		\multicolumn{1}{|c||}{TX/RX antenna gain (dBi)} &   \multicolumn{1}{c|}{$10$ } & \multicolumn{1}{c|}{$24$ } & \multicolumn{1}{c|}{$15$} \\ 
		\hhline{-|-|--}
		
		\multicolumn{1}{|c||}{TX/RX antenna 3 dB beam width in E plane, ($\deg$)} & \multicolumn{1}{c|}{$54.2$}  & \multicolumn{1}{c|}{$9.16 $} & \multicolumn{1}{c|}{$32 $} \\ 
		\hhline{-|-|--}
		
			\multicolumn{1}{|c||}{TX/RX antenna 3 dB beam width in H plane, ($\deg$)} & \multicolumn{1}{c|}{$54.4$}  & \multicolumn{1}{c|}{$9 $} & \multicolumn{1}{c|}{} \\ 
		\hhline{-|-|--}
		
		\multicolumn{1}{|c||}{Antenna diameter, A (cm)} & \multicolumn{1}{c|}{$1.3 $} & \multicolumn{1}{c|}{$3.77$} & \multicolumn{1}{c|}{$ 0.8$} \\ 
		\hhline{-|-|--}
		
		\multicolumn{1}{|c||}{Antenna far-field for $dmax$ (cm)} & \multicolumn{1}{c|}{$3.1$} & \multicolumn{1}{c|}{$6.9$} & \multicolumn{1}{c|}{$4$} \\ 
		\hhline{-|-|--}		

		\multicolumn{1}{|c||}{Projection diameter, D for $dmax$ (cm)} & \multicolumn{1}{c|}{$30.7$} & \multicolumn{1}{c|}{$4.72$} & \multicolumn{1}{c|}{$17.2 $} \\ 
		\hhline{-|-|--}

	\end{tabular} 
\end{table*}


 To the best of authors' knowledge, this article presents for the first time specific attenuation data for building materials at $91$ GHz. To collect a set of comparative measurement data, we also measured  specific attenuation at $28$ and $73$ GHz for the same building materials.
This paper presents a unique set of results on attenuation measurements conducted by three universities\textemdash University of South Carolina (USC), Boise State University (BSU) and North Carolina State University (NCSU). We performed attenuation measurements for building materials using both wideband and narrowband measurements at three mmwave frequency bands in a laboratory environment. Our measurement results can aid designers in accounting for specific building material losses in link budget calculations for future mmWave communication systems.

This paper is structured as follows: Section~\ref{sec:method} briefly summarizes the measurement procedure we used. Section~\ref{sec:setup} describes experimental equipment setups for the attenuation measurements in the three bands, separately conducted by the three teams. In Section~\ref{sec:results}, we provide results collected from the measurement campaigns for the three frequency bands. Finally, Section~\ref{sec:conclusion} concludes this paper.

\section{Measurement Procedure}
\label{sec:method}

Measurements at the three frequencies were done by three different research groups: 28 GHz was used by NC State University, $73$ GHz by Boise State University, and $91$ GHz by the University of South Carolina. Before beginning measurements we did some simple geometric calculations to ensure that the main lobe antenna projection on the material does not exceed the material dimensions. Otherwise  diffracted and/or multipath components can reach the receiver and degrade material attenuation measurement accuracy. The materials and their dimensions are listed in Table~\ref{tab:material_model}. 

Fig.~\ref{setup} and Fig.~\ref{setup2} show our geometry for properly choosing dimensions for the materials. This allows for the antenna main beam projection area plus a guard area, where the main beam projection diameter is, 
\begin{equation}D=2d_1\sin(\alpha)~,\label{eq}
\end{equation}
where $\alpha$ is half antenna beam width and $d_1$ is the  distance between antenna aperture and the material. Note that \eqref{eq} should be calculated for $d_max$ presented in Table \ref{tab:hardware_spe}. In this campaign we used a guard annulus width $g=r/2$ where $r$ is the projected circle radius on the material and equal to $r=D/2$. The material height and width was  chosen to provide additional margin (material dimension $\gg r+g$).
As also discussed in \cite{different_concrete}, diffraction from the sample edges can impose significant errors, hence we used the ``guard region" to minimize this effect.  Identical materials were purchased from one retailer in the US for each of the three research groups. This enables comparing results for identical materials at the different frequencies. Distances to the materials were also chosen so that the Tx and Rx antennas were located in the far-field of both Tx and Rx antennas, i.e., 
\begin{equation}d_1 > 2A^2/\lambda~,\label{eq_farfield}
\end{equation}
where $\lambda$ is signal wavelength and $A$ is horn antenna aperture dimension. These distances are listed in Table \ref{tab:hardware_spe}.

Based on dimensions in Table \ref{tab:hardware_spe}, the 28 GHz antenna yielded the largest projection diameter of 30.7 $ cm$ and the 73 GHz antenna had the largest far field distance. Based on these values we chose measurement distances.

 \begin{figure}[!t]
\centerline{\includegraphics[width=\columnwidth]{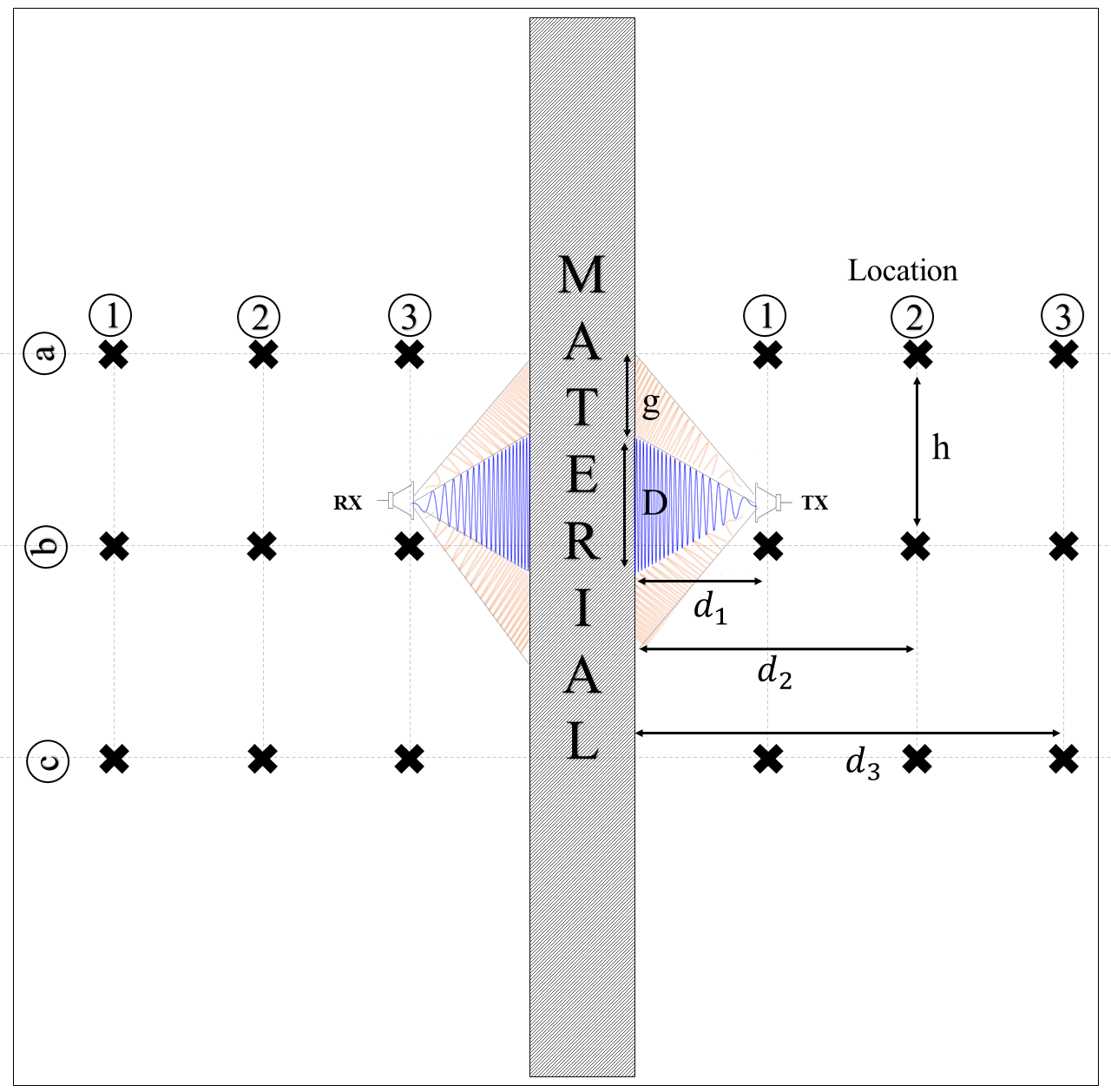}}
\caption{Measurement Procedure}
\label{location}
\end{figure}
 \begin{figure*}[!t]
\centerline{\includegraphics[width=1\textwidth]{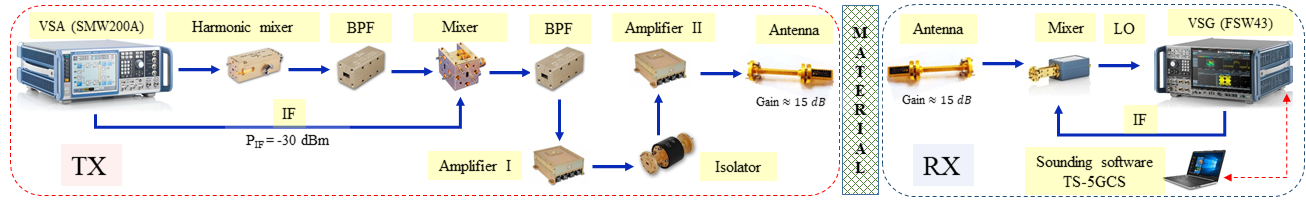}}
\caption{$91$ GHz setup for attenuation measurement. }
\label{fig_91_setup}
\end{figure*}
 
 \begin{figure*}[!t]
\centerline{\includegraphics[width=1\textwidth]{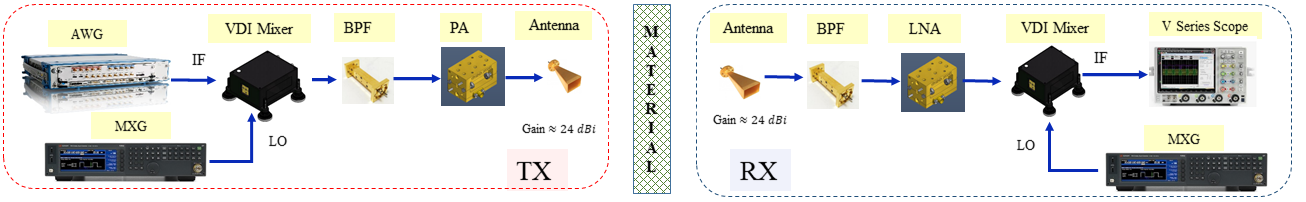}}
\caption{$73$ GHz setup for attenuation measurement. }
\label{fig_73_setup}
\end{figure*}
 \begin{figure*}[!t]
\centerline{\includegraphics[width=1\textwidth]{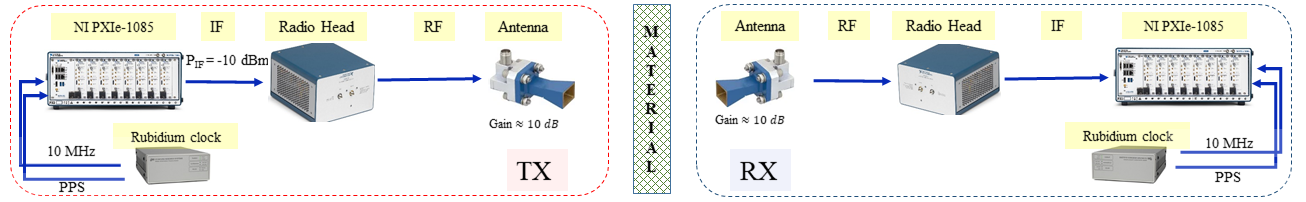}}
\caption{$28$ GHz setup for attenuation measurements. }
\label{fig_28_setup}
\end{figure*}

Fig.~\ref{location} illustrates the nine locations at which we took material attenuation measurements. Designated distances $d_1$, $d_2$, $d_3$ and $h$ are 10, 20, 30 and 4 cm, respectively. The reason for multiple locations is to minimize error due to antenna beam misalignment and material inhomogeneity. For the narrowband signal tests, this also enables averaging out any residual small-scale fading effects. Attenuations recorded for each location were averaged over all nine locations.  Along with knowledge of transmit power, estimated path loss is calculated by ~(\ref{eq1})  \cite{rappaport1996wireless},

\begin{align}
\label{eq1}  
PL(d)(\text{dB}) =  P_t(\text{dBm})-P_{r}(d)(\text{dBm})  
\nonumber
        \\
         + G_t(\text{dB}) + G_r(\text{dB}) ,
\end{align}
where $d $ is the TX-RX separation distance, $P_t$ is the transmit power, $G_t$ and  $G_r$ are the maximum gains of transmitter and receiver antennas, respectively, and $P_{r}(d)$ is the measured received power at a distance of $d$. The attenuation $L$ is computed as the difference in dB between average received power with the test material and the received power in unobstructed free-space with the same Tx-Rx separation distance. Mathematically $L(dB)$ can be calculated at

\begin{align}
\label{eq2}
L(dB) = PL^{unobs.}(d) - PL^{MUT}(d) ,
\end{align}
where, $PL^{MUT}(d)$ is the path loss for the material under test (MUT), and $PL^{unobs.}(d)$ is the path loss in the unobstructed free-space measurement.

Most building materials are affected by the surrounding environment parameters, i.e. humidity and temperature. All measurements in this paper were done at room temperature (approximately $22 C\degree$ ) and humidity between $20$ to $60\%$.


\begin{table*}[!t]
	\centering
	\caption{Average specific attenuation with standard variation at $28$, $73$ and $91$ GHz in three locations for wideband measurements. Only narrowband results are enlisted for concrete block at $91$ and $73$ GHz because of the larger attenuation. Variable $\sigma$ is the standard deviation of the attenuation over
		all locations for the same material.}
	\label{tab:resulst}
	\begin{tabular}{V{5}c||c||c||c||c||cV{5}}
		\hlineB{5}
		Frequency/wavelength                & Material          & Thickness, mm &   Mean attenuation, dB/cm          & Standard deviation dB$\sigma$ & Total attenuation, dB         \\ \hlineB{3}
		
		\multirow{5}{*}{$91$ GHz} & clear glass & $1.8$   & $18.79$ & $6.39$& $3.38$ \\ \cline{2-6} 
		                          & plywood & $11.2$ &$11.76$ & $4.2$& $13.17$ \\ \cline{2-6} 
		                          & drywall &  $13.1$ &  $1.97$ & $1.26$& $2.58$ \\ \cline{2-6}
		                          & cinder block &  $195$  & $2.13$ & $0.22 $ & $41.53 $\\ \cline{2-6}
		                          & acoustic ceiling tiles & $11.3$&  $1.0$ & $1.14$& $1.13$  \\ \hlineB{3}
		
		\multirow{5}{*}{$73$ GHz} & clear glass & $1.8$&    $14.37$ & $5.76$& $2.58$ \\ \cline{2-6}
		                          & plywood&   $11.2$&  $7.14$ & $0.81$& $7.99$ \\ \cline{2-6}
		                          & drywall &  $13.1$ &   $1.45$ & $0.73$& $1.89$ \\ \cline{2-6}
		                          & cinder block &  $195$ &  $1.9$ & $0.19$ & $37.05$ \\ \cline{2-6}
		                          & acoustic ceiling tiles &   $11.3$&   $0.93$ & $0.52$& $1.05$ \\ \hlineB{3}
		
		\multirow{5}{*}{$28$ GHz} &  clear glass &  $1.8$   & $4.38$ & $0.18$& $0.78$ \\  \cline{2-6}
		                          & plywood &   $11.2$ & $5.09$ & $1.28$& $5.7$ \\ \cline{2-6} 
		                          & drywall &   $13.1$ & $0.94$ & $0.23$& $1.23$\\ \cline{2-6}
		                          & cinder block &  $195$ & $1.03$& $0.18$& $20.08$ \\ \cline{2-6}
		                          & acoustic ceiling tiles &  $11.3$ &   $0.44$ & $0.06$& $0.49$ \\ \hlineB{3}
	\end{tabular}
\end{table*}


 \section{Measurement Setup}
 \label{sec:setup}
 In this section, we briefly explain the measurement setup for each of the three bands\textemdash $91$, $73$ and $28$ GHz, as conducted by the three different university teams.
\subsection{Frequency Band $91$ GHz }
\label{sec:setup3}
 The 91 GHz measurement setup is shown in Fig.~\ref{fig_91_setup} where a Rohde \& Schwarz {(R\&S) vector signal generator (VSG) SMW200A} and {signal and spectrum analyzer (SSA)} FSW43 act as a transmitter and receiver,  respectively.  The VSG specification manual \cite{VSG_SPEC} claims $<$0.9 dB output level inaccuracy for temperatures ranging from +18 to +33 {C\degree}  for 11 GHz. For the SSA, level measurement uncertainty (accuracy) is reported in \cite{VSA_SPEC}, which for the IF output frequency of our harmonic mixer (7-13 GHz) is 1.5 dB with standard deviation of 0.5 dB in +20 to +30 {C\degree} temperatures. These inaccuracies should not affect our results since in (\ref{eq2}) subtraction will remove them. A Quinstar harmonic mixer up converts the sinusoidal LO 11 GHz VSG signal to 88 GHz. This is then band pass filtered and mixed with a 3 GHz intermediate frequency (IF) signal to generate a signal centered at 91 GHz. Remaining components (all Quinstar) include a bandpass filter, amplifiers and an isolator, as depicted in Fig. \ref{fig_91_setup}. The antenna was a SAGE model 1532-10-S2 with 15 dBi gain and 32 degrees beam width (symmetric in azimuth and elevation). At the receiver, another R\&S external mixer downconverts the signal to an appropriate frequency range for SSA analysis.

 For narrowband measurements our IF input signal was a single tone. For wideband measurements we used the R\&S TS-5GCS channel sounding software. This software has multiple signals that can be transmitted and received; we employed a filtered chirp with flat spectrum and very sharp rolloff, with bandwidth 500 MHz. The chirp detection and post processing at the receiver is explained in \cite{nozhan}.
 
\subsection{Frequency Band $73$ GHz }
\label{sec:equipment_73}
For the $73$ GHz attenuation measurements, we used a Keysight M8190A  wideband arbitrary waveform generator (AWG) as the transmitter, and a Keysight DSA V084 oscilloscope as the receiver~\cite{mahfuza_vtc,mahfuza2018,mahfuza_globcom19}. The waveform generator and scope were connected and synchronized using a $10$ MHz signal. The IF signal at $4$ GHz from the AWG was mixed with a local oscillator to reach $73$ GHz. The LO frequency was set at $38.5$ GHz, which is multiplied by two. Subsequently, we used a band pass filter to remove any unwanted signals. Next, a power amplifier with a gain of $20$ dB was used before the TX antenna. The horn antenna has a gain of $24$ dBi and a $3$ dB beamwidth of $9^{\circ}$ and  $9.16^{\circ}$ in azimuth and elevation planes, respectively. 

At the receiver, a scope, a downconverter, a local-oscillator, a low noise amplifier, and a band pass filter were employed for receiving the transmitted IF signal at $4$ GHz. An identical horn antenna was used at the Rx. After mixing down, the received signal was fed to the scope, where received signal strength was measured using Keysight 89600 VSA software. Fig.~\ref{fig_73_setup} shows the measurement test setup at $73$ GHz, where both the Tx and the Rx were placed atop instrument carts ~\cite{mahfuza_globcom19}. We generated a chirp and a CW signal for the wideband and narrowband measurements, respectively. A wideband signal correlation time domain channel sounding approach was employed for the  wideband measurements. Detailed hardware parameters are provided in Table ~\ref{tab:hardware_spe}.

\subsection{Frequency Band $28$ GHz }
\label{sec:setup_28}
For $28$ GHz attenuation measurements, we used the National Instruments~(NI) channel sounder hardware at $28$~GHz~\cite{NImmwave}. This sounder has been used in our previous measurements \cite{wahab_indoor,wahab_outdoor}, and consists of NI PXIe-1085 TX/RX chassis and 28~GHz TX/RX mmWave radio heads from NI. The 10 MHz and pulse per second~(PPS) signals generated by an FS725~Rubidium~(Rb) clock~\cite{SRS} were connected to PXIe~6674T modules at both TX and RX. The common 10 MHz signal was used to generate the required local oscillator (LO) signals and the PPS signal was used to trigger the transmission and reception of the sounding waveform. 

The sounder software was based on LabVIEW, and a Zadoff-Chu (ZC) sequence of length 2048 was periodically transmitted to measure the channel. The ZC sequence was filtered by a root-raised-cosine (RRC) filter, and the generated samples were uploaded to PXIe-7902 FPGA. These samples were sent to PXIe-3610 digital-to-analog converter (DAC) with a sampling rate of $f_s=3.072$~GS/s. The PXIe-3620 module up-converted the base-band signal by multiplying it with a signal 3 times the first LO signal at $3.52$ GHz to reach an IF of $10.56$ GHz. The 28~GHz mmWave radio head further up-converted the IF signal by multiplying the second LO at $4.82$ GHz by 8. At the Rx, the 28~GHz mmWave radio head down-converted the RF signal to IF and was down-converted again in the PXIE-3620 module to base-band. The PXIe-3630 analog-to-digital converter (ADC) module sampled the base-band analog signal with a sampling rate of $f_s=3.072$~GS/s. The correlation and averaging operations were performed in PXIe-7902 FPGA operation, and the complex CIR samples were sent to the PXIe-8880 host PC for further processing. Before the measurement, calibration was performed to eliminate the channel distortion caused by the non-idealities of the hardware. The directional horn antennas we used have specification as shown in Table~\ref{tab:hardware_spe}: the antennas had 10~dBi gains, and $54.2^{\circ}$ and $54.4^{\circ}$ beam-widths in the elevation and azimuth planes, respectively.
Fig.~\ref{fig_28_setup} shows our equipment setup. The Tx and Rx radio heads were fixed to 2 boxes and the materials to be tested were placed between them. 
\begin{figure}[t!]
   \centerline{\includegraphics[width=\columnwidth]{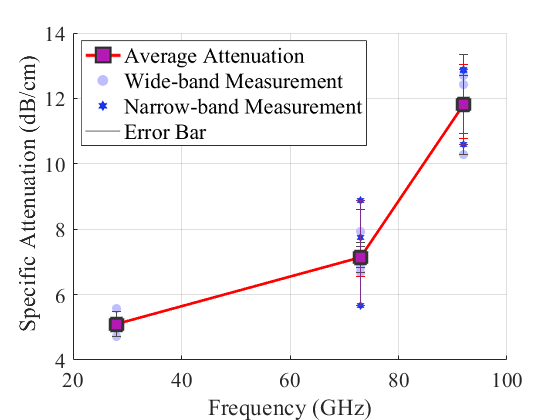}}
   \caption{Plywood attenuation vs. frequency.
 }
   \label{fre_attn_plywood}
\end{figure}

\begin{figure}[!t]
\centerline{\includegraphics[width=\columnwidth]{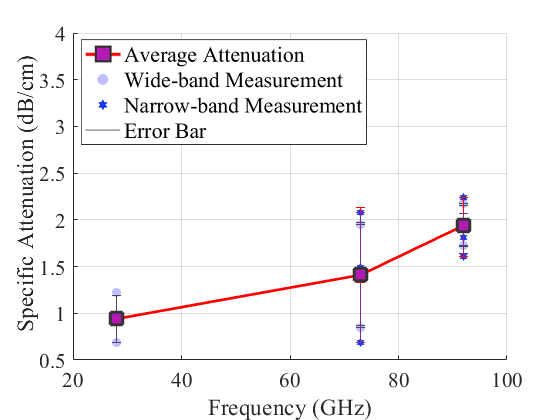}}
\caption{Drywall attenuation vs. frequency.
 }
\label{fre_attn_drywall}
\end{figure}

\section{Measurement Results}
\label{sec:results}

Table~\ref{tab:resulst} presents a summary of the average specific attenuation of building materials at $28$, $73$ GHz and $91$ GHz for wideband measurements. The standard deviations of attenuation across all the locations of each material are also provided as well as total material attenuation in dB. As presented in Table \ref{tab:resulst}, the clear glass has the highest attenuation among all standard materials at $73$ and $91$ GHz. For example, the largest average specific attenuation of clear glass was measured to be $18.79$ dB/cm and $14.37$ dB/cm for the $91$ and $73$ GHz frequencies, respectively. The lowest values of specific attenuation were found for the acoustic ceiling panels, and were $1.0$ and $0.93$ dB/cm at $91$ and $73$ GHz, respectively. For the $91$ GHz and $73$ GHz frequency bands, the largest and smallest standard deviations were found for clear glass and acoustic ceiling tiles, respectively. For $28$ GHz, the largest standard deviation was measured as $1.28$ dB for plywood, and the lowest was obtained as $0.06$ dB for acoustic ceiling tile.

As noted, wide band measurement results were obtained via a frequency modulated signal (chirp) sweeping across the bandwidth, with the receiver using either a matched filter or heterodyne detector~\cite{nozhan}. The wideband signals can also show any attenuation variation within the bandwidth. Wide band and narrow band measurement results (at $73$ and $91$ GHz) show a very close agreement: we have determined that the small difference is attributable to the out of band energy of the wide band signal that was not fully accounted for in our systems.

\begin{figure}[!t]
\centerline{\includegraphics[width=\columnwidth]{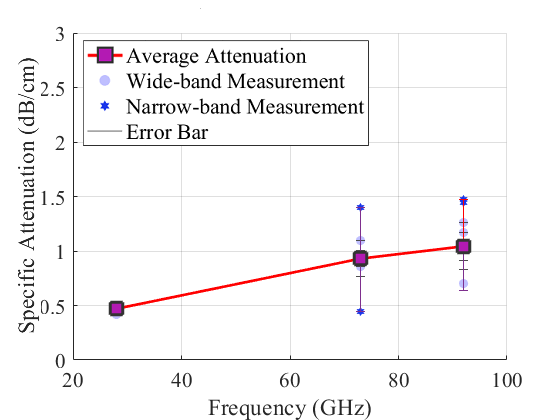}}
\caption{Acoustic ceiling tail attenuation vs. frequency.}
\label{fre_attn_ceil}
\end{figure}

\begin{figure}[!t]
\centerline{\includegraphics[width=\columnwidth]{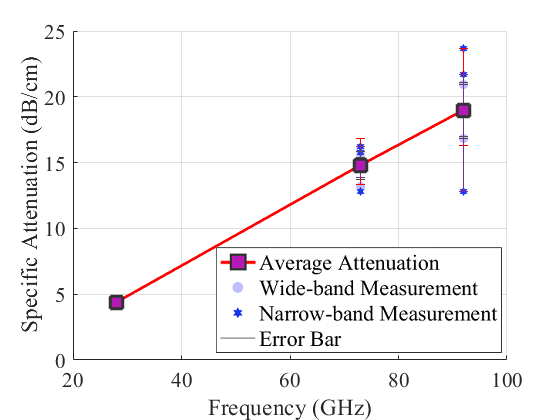}}
\caption{Clear glass attenuation vs. frequency.
 }
\label{fre_attn_glass}
\end{figure}

\begin{figure}[!t]
\centerline{\includegraphics[width=\columnwidth]{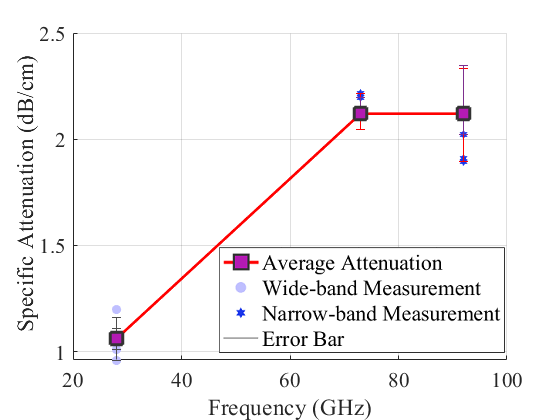}}
\caption{Cinder blocks attenuation vs. frequency.
 }
\label{fre_attn_concrete}
\end{figure}

Figures ~\ref{fre_attn_plywood} to \ref{fre_attn_concrete} show the measured results for the different investigated materials. Both narrow-band and wide-band measurement results are presented, along with the average of all data points. Bars show the range of measured data for the $9$ locations (averaged along rows $a$ to $c$ in Fig. \ref{location}).
Fig.~\ref{fre_attn_plywood} shows plywood attenuation for the three frequencies. Plywood is a composite sheet material made from  several thin layers of wood that are glued together with adjacent layers having their wood grain rotated up to 90 degrees from one another. The selected plywood wood grain is pine and is commonly used as a panel in outside building structures, walls, and partitions.  

 \begin{figure}[!t]
\centerline{\includegraphics[width=\columnwidth]{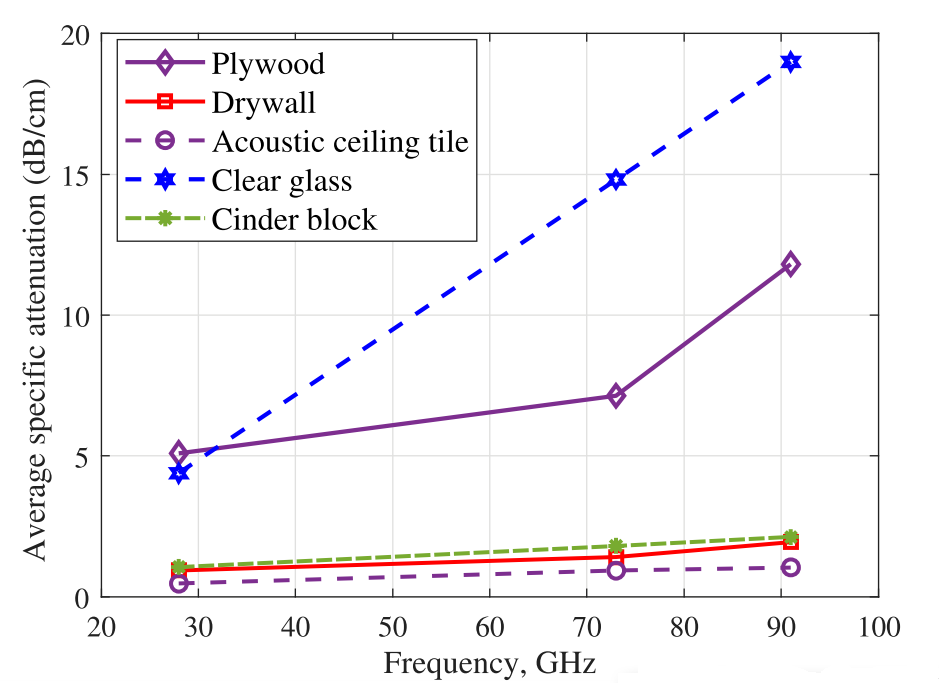}}
\caption{Wideband average attenuation versus frequency for all test materials
 }
\label{fre_attn}
\end{figure}
\textcolor{black}{ Fig.~\ref{fre_attn_drywall} shows the frequency dependent attenuation for drywall panels.  These drywall panels (or wallboard) are made of calcium sulfate dihydrate and some additional materials such as paper and fiberglass. These panels are very commonly used in building interior walls and ceilings. }
 
Fig.~\ref{fre_attn_ceil} shows attenuation for acoustic ceiling tiles for the different frequencies. Acoustic ceiling panels are designed to block, absorb, and diffuse sound. PVC, nylon and mineral fibers are typically used in acoustic ceiling tile construction.

Fig. \ref{fre_attn_glass} shows the clear glass attenuation for our frequency values. Clear glass is not only used in many building widows but also for exterior building walls, either within, or instead of concrete walls

Fig. ~\ref{fre_attn_concrete} shows attenuation for a set of cinder blocks that are often used in building walls. We note that these cinder blocks have two air holes inside and are not completely solid. Therefore, our mmWave signals can travel through these holes and attenuation differs based on where the antennas are located. As we can see in the measured results, attenuation is not consistent and varies significantly due to antenna location. As mentioned in \cite{different_concrete}, although these blocks were purchased from the same retailer in the United States, their water-to-cement ratio (and hence attenuation) may differ among our locations. 
Finally, Fig. \ref{fre_attn}  presents average specific attenuation for all the materials in one plot. We  observe that clear glass has the largest average specific attenuation, and acoustic ceiling tile has the smallest average specific attenuation. 

\section{Conclusion}
\label{sec:conclusion}
\textcolor{black}{In this paper, we reported on building material attenuation in three different mmWave frequency bands ($28$ GHz, $73$ GHz and $91$ GHz). Standard building materials were ordered from the same retailer to have as close to identical materials as possible for comparison. These materials were clear glass, cinder blocks, plywood, drywall, and acoustic ceiling tiles. The measurement distances were calculated to ensure minimal signal distortion  from other objects in the measurement environment, i.e., we strove to minimize effects from reflections, diffraction and multi-path components. Nine measurement locations were used for both narrow band and wideband signals to reduce the effects of misalignment error and material inhomogeneity. Data was presented to show the range of measured variation caused by these effects. The largest specific attenuation at $91$ GHz and $73$ GHz was for clear glass, with values 18.79 and 14.37 dB/cm, respectively, and 5.05 dB/cm  for plywood at $28$ GHz. The smallest specific attenuation, in all three bands, was for acoustic ceiling tiles with values 1.0, 0.93, and 0.44 dB/cm at 91, 73, and 28 GHz, respectively.} For future work, we plan to add more standard materials to our database. Our results should help designers to  account for the attenuations of these building materials in link calculations.

\bibliographystyle{IEEEtran}

\bibliography{IEEEabrv,ref}



%
\end{document}